# Mechanisms for a Spring Peak in East Asian Cyclone Activity

Satoru Okajima,[a] Hisashi Nakamura,[a] Akira Kuwano-Yoshida,[b] Rhys Parfitt[c]

[a] *Research Center for Advanced Science and Technology, The University of Tokyo, Tokyo, Japan*

[b] *Disaster Prevention Research Institute, Kyoto University, Kyoto, Japan*

[c] *Florida State University, Tallahassee, Florida*

ABSTRACT

The frequency of extratropical cyclones in East Asia, including those traveling along the Kuroshio off the south coast of Japan, maximizes climatologically in spring in harmony with local enhancement of precipitation. The springtime cyclone activity is of great socioeconomic importance for East Asian countries. However, mechanisms for the spring peak in the East Asian cyclone activity have been poorly understood. This study aims to unravel the mechanisms, focusing particularly on favorable conditions for relevant cyclogenesis. Through a composite analysis based on atmospheric reanalysis data, we show that cyclogenesis enhanced around the East China Sea under anomalously strengthened cyclonic wind shear and temperature gradient, in addition to enhanced moisture flux from the south, is important for the spring peak in the cyclone activity in East Asia. In spring, climatologically strengthened cyclonic shear north of the low-level jet axis and associated frequent atmospheric frontogenesis in southern China and the East China Sea serve as favorable background conditions for low-level cyclogenesis. We also demonstrate that climatologically enhanced diabatic heating around East Asia is pivotal in strengthening of the low-level jet through a set of linear baroclinic model experiments. Our findings suggest the importance of the seasonal evolution of diabatic heating in East Asia for that of the climate system around East Asia from winter to spring, encompassing the spring peak in the cyclone activity and climatological precipitation.

SIGNIFICANCE STATEMENT

This study aims to elucidate mechanisms for a spring peak in the frequency of low-pressure systems in East Asia. We show that the climatological development of the low-level jet stream in southern China and the East China Sea in spring is important for the more frequent generation of those low-pressure systems. We also demonstrate that climatological increase in diabatic heating in East Asia from winter to spring strengthens the low-level jet in southern China and the East China Sea. Our findings suggest the importance of the seasonal evolution of diabatic heating in East Asia for the seasonal evolution of the East Asian climate system from winter to spring, including the spring peak in low-pressure system frequency, climatological rainfall.

# 1. Introduction

Giving rise to day-to-day weather variations and precipitation in the extratropics, synoptic-scale migratory cyclones transport heat poleward and they are thus one of the fundamental components for the climate system. The activity of those extratropical cyclones (ETCs) in the Northern Hemisphere is overall stronger in the cold season than in the warm season (e.g., Hoskins and Hodges 2002). This tendency is consistent with the seasonality of the maximum growth rate of baroclinic eddies, although their activity does not necessarily correspond to Eulerian eddy statistics (Okajima et al. 2021).

In East Asia (EA), ETC tracks in the cold season are climatologically concentrated along multiple typical paths (Asai et al. 1988; Chen et al. 1991; Adachi and Kimura 2007). The southernmost path starts from eastern China, extending over the East China Sea (ECS) and off the south coast of Japan along the Kuroshio before reaching into the Kuroshio Extension (KE) east of Japan. Unlike the other typical paths, the southernmost path is characterized by a spring peak in the climatological ETC frequency. This distinct seasonality of the ETC activity has been documented on the basis of weather charts (Asai et al. 1988; Chen et al. 1991) and objective cyclone tracking algorithms (Adachi and Kimura 2007; Wang et al. 2009; Lee et al. 2020) applied to atmospheric reanalysis data.

In the EA region spanning from southern China to Japan, climatological-mean precipitation at many stations exhibits temporal enhancement or even a peak in March (Fig. 1). The spring peak in climatological-mean precipitation corresponds to that in ETC frequency, thereby suggesting the linkage between spring ETC activity in EA and spring rainy season in southern China (Tian and Yasunari 1998; Qian et al. 2002; LinHo et al. 2008). Cho et al. (2018) reported that the springtime ETC activity in EA has weakened from the late 1970s into the 2000s, concurrent with a decreasing trend of springtime rainfall in southern China. Spring rainfall is important for crop production and plant community assembly (Hatfield et al. 2011; Larson et al. 2021). Additionally, ETCs moving along the south coast of Japan can cause hazardous heavy snowfall in densely populated areas in Japan (Honda et al. 2016). They also tend to cause low variable renewable energy generation through cloudy and low-wind conditions (Ohba et al. 2023). Therefore, springtime ETC activity is of great socioeconomic importance for the EA countries, whose large populations are thriving with agriculture and demand a large amount of electricity.

File generated with AMS Word template 2.0

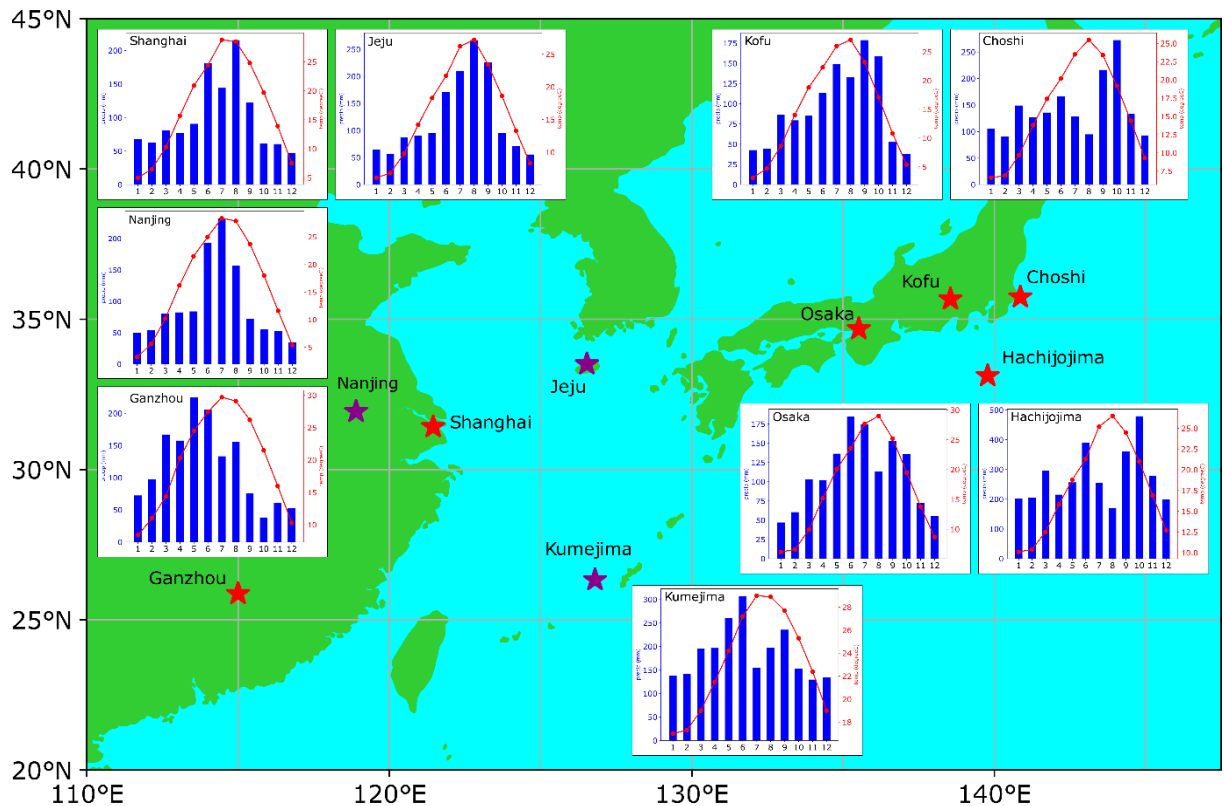

Fig. 1. Seasonal evolution of climatological-mean monthly temperature (red lines; °C) and precipitation (blue bars; mm) at locations as indicated by stars on the map, based on CLIMAT data provided by the Japan Meteorological Agency (JMA). Climatologies are defined as averages for the period 1991−2020. Red and purple stars denote stations where climatological precipitation exhibits a peak in March and exhibits just enhancement in March, respectively.

Nevertheless, mechanisms for the spring peak in the ETC activity in EA are still poorly understood. This contrasts with the effort that has recently been made to investigate individual extreme snowfall events associated with ETCs on the south coast of Japan (Takano 2002; Honda et al. 2016) and characteristics of explosive cyclones (Yoshida and Asuma 2004; Hirata et al. 2015; Kuwano-Yoshida et al. 2022). The dynamical understanding of the climatological springtime peak of ETC activity will be useful for investigating its interannual to decadal-scale variability, predictability, and future changes.

The present study thus aims at unraveling the mechanisms for the spring peak in the ETC activity in EA, focusing particularly on favorable conditions for relevant cyclogenesis. This paper is organized as follows; Section 2 explains data and analysis methods used in this study. Section 3 describes the seasonality of ETC activity in EA. Section 4 shows typical conditions for ETCs and cyclogeneses around the ECS. Section 5 discusses mechanisms for the springtime development of the low-level jet around the ECS. Section 6 offers a summary and discussions.

## 2. Data and methods

*a. Observational data*

We analyze 6-hourly global fields of atmospheric variables, including geopotential height, air temperature, wind velocities, specific humidity, and diabatic heating rates in pressure coordinates, in addition to sea-level pressure (SLP), obtained from the Japanese 55-year atmospheric reanalysis (JRA-55) produced by the Japan Meteorological Agency (JMA) (Kobayashi et al. 2015; Harada et al. 2016) for the period 1979−2018. The JRA-55 has been constructed with a four-dimensional variational data assimilation (4D-Var) system with TL319 horizontal resolution (equivalent to 55-km resolution).

Variables at selected pressure levels are available on a 1.25°×1.25° grid. At each grid point, Eulerian eddy statistics are calculated from fluctuations of a given variable with synoptic-scale transient eddies that are extracted locally from the 6-hourly reanalysis data as its deviations from their low-pass-filtered fields through a 121-point Lanczos filter with a cutoff period of 8 days. Plots showing seasonal evolutions (as in Fig. 2) and composite maps (as in Fig. 5) are produced after applying a 31-day running mean to the daily climatology and composited fields. A climatological-mean field is based on the period of 1979−2018 unless otherwise specified.

We also analyze monthly precipitation from the Global Precipitation Climatology Project (GPCP) version 3.2 (Huffman et al. 2023) for the period 1983−2017 in Fig. 2.

*b. Cyclone tracking*

In this study, tracks of surface migratory ETCs are objectively identified by the algorithm used by Kuwano-Yoshida et al. (2022) and Okajima et al. (2023). Assessment of the performance of the tracking algorithm is provided by Okajima et al. (2023). The algorithm is based on a 6-hourly unfiltered SLP field. The use of SLP helps avoid the influence of shear vorticity near the low-level jet. The algorithm requires a cyclone track to exist not less than 24 hours (four time steps). In this study, a cyclogenesis event is identified as the first time step of a given cyclone track[1].

## 3. Seasonality of cyclone activity in East Asia

In spring (FMA (February−April)-mean), a distinct band of high ETC density is climatologically located off the south coast of Japan (Fig. 2a), which starts from the ECS and merges into the North Pacific (NP) storm-track core to the east of Japan. The peak of ETC density roughly follows the Kuroshio and KE, and it is located just north of the low-level jet axis (Fig. 2a). In winter (DJF (December−February)-mean), the local peak along the Kuroshio is substantially weaker than its springtime counterpart (Fig. 2b). The low-level jet south of Japan and around the ECS is less distinct in winter than in spring, despite the stronger low-level jet core over the NP. Figure 2c depicts a distinct enhancement in the springtime ETC density along the Kuroshio at ~32°N north of the modest low-level jet axis at ~30°N, peaking in mid-to-late March. The prominent spring peak of ETC frequency can be also seen for 1958-2022 (Supplementary Fig. S1). The spring peak contrasts with the midwinter (or early winter) peak in ETC density in the northern/central portions of the Japan Sea, concurrent with the maximized low-level jet speed. The intensity of ETCs measured by $\nabla^2$SLP exhibits no distinct spring peak along the Kuroshio (Supplementary Fig. S2). The cyclone intensity in EA south of ~30°N is likely to be influenced by tropical cyclones in autumn but unlikely in winter and spring.

---

[1] This means that a cyclone is not detected until it reaches decent intensity close to the synoptic scale. A weak, meso-scale cyclone in southern China in the incipient stage of ETC advancing along the south coast of Japan in Takano (2002) is unlikely to be identified.

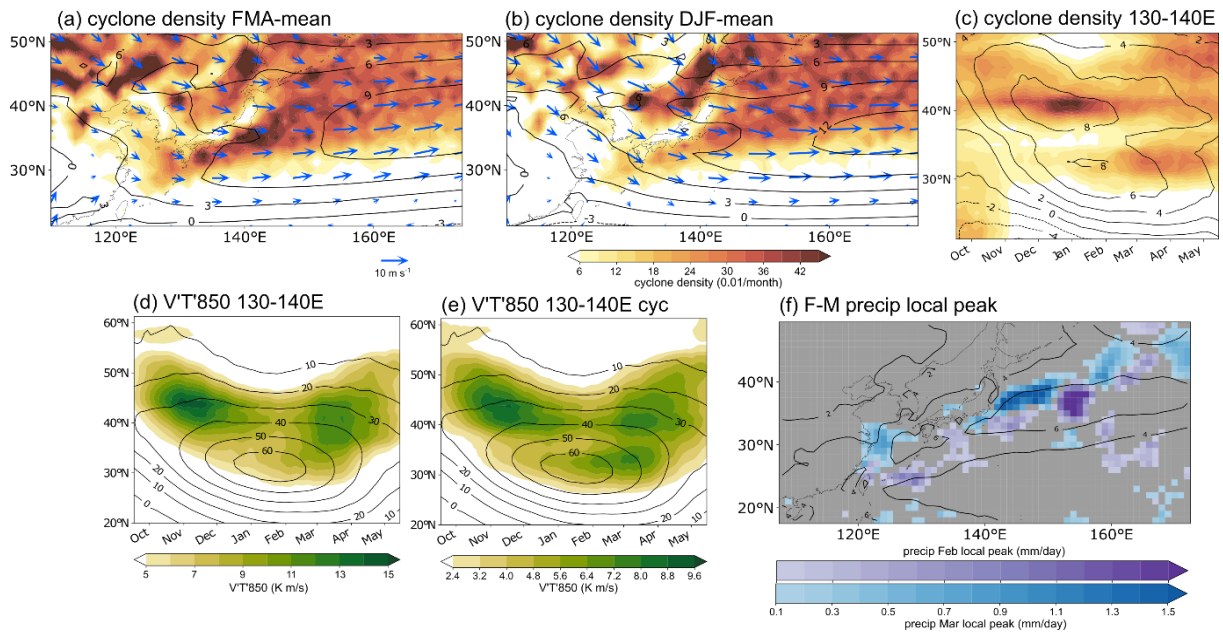

Fig. 2. (a) Climatological-mean springtime (FMA-mean) cyclone density (colors, 0.01/month). Contours denote climatological-mean westerly wind speed at 850-hPa ($U_{850}$; m/s). Arrows signify climatological-mean horizontal wind as indicated. (b) Same as in (a), but for wintertime (DJF-mean) climatologies. (c) Climatological seasonality in ETC density (colors, $10^{-4}$/6hr) averaged for 130°−140°E as a function of latitude. Contours denote the corresponding seasonality of $U_{850}$ averaged for 130°−140°E. A tick mark on the abscissa in b-c represents the first day of a given calendar month. (d) Same as in (c), but for $V'T'_{850}$ (color, K m/s). Contours denote the corresponding seasonality of $U_{300}$. (e) Same as in (d), but for the cyclonic contribution to $V'T'_{850}$ with a curvature threshold of zero. (f) Amplitude of a peak in climatological-mean monthly precipitation (mm/day) in February (purple) and March (blue) measured by the comparison with the greater climatological precipitation of the adjacent calendar months, based on GPCP v3.2. Only grid points with monthly precipitation more than 0.1 mm/day are plotted. Contours denote climatological-mean FM-mean precipitation (mm/day).

The spring peak in ETC density corresponds to southward expansion of springtime low-level storm-track activity measured by poleward eddy heat flux ($V'T'_{850}$) in the NP storm-track entrance region (Fig. 2d). The springtime enhancement corresponds to the second peak after the midwinter minimum of the NP storm-track activity despite the midwinter maximum in the Pacific jet intensity (Nakamura 1992; Okajima et al. 2024). As in Fig. 2e, the southward expansion of springtime low-level storm-track activity is more evident in the cyclonic contribution to $V'T'_{850}$ based on the methodology by Okajima et al. (2021). Correspondingly, the springtime enhancement of both cyclone density and low-level storm-track activity leads to a local peak in precipitation on the east coast of China and the south coast of Japan as well as around the KE in February and March (Fig. 2f), which is consistent with the results from station data (Fig. 1).

To illustrate a typical springtime ETC migrating along Kuroshio south of Japan, we focus on a domain [135°−141.25°E, 28.75°−36.25°N] (hereafter, called "SEJ"; southeastern Japan), which includes local maxima of ETC density (Fig. 2a). We also focus on composite maps and climatological-mean fields on 19th Mar (hereafter called "springtime" composite and climatology, respectively) after a 31-day running-mean is applied (see Section 2). This calendar day corresponds to the second peak after the midwinter minimum of the NP storm-track activity (Fig. 2d; Okajima et al. 2023). A composited springtime ETC within the SEJ domain well captures a typical structure of the so-called "south-coast cyclone" (e.g., Nakamura et al. 2012; Fig. 3a), which is referred to as a "Kuroshio cyclone" in this study. The composited ETC, whose center is located just to the south of the Kuroshio axis, accompanies a precipitation maximum slightly to the northeast of its center. The Kuroshio cyclones passing through the SEJ domain well reproduce the local maxima in the total climatological-mean springtime ETC density extending northeastward from the ECS into the KE (Figs. 2a and Fig. 3b). As shown in Fig. 3c, those Kuroshio cyclones are generated mainly over the ECS with secondary contributions from cyclogenesis to the south of eastern and western Japan. To quantify the relative importance of cyclogenesis over these three domains for the spring peak in the Kuroshio cyclone density passing through the SEJ domain, we calculate the ETC density averaged over the SEJ domain based only on ETCs generated within each of the ECS [115°−130°E, 25°−35°N], southwestern Japan (SWJ) [130°−135°E, 28.75°−35°N], and SEJ domains, as indicated in Fig. 3c. As evident in Fig. 3d, the spring peak in the Kuroshio cyclone density over the SEJ domain is accounted for mostly by ETCs generated within the ECS domain, although the cyclogenesis in the SWJ and SEJ domains also peaks in April.

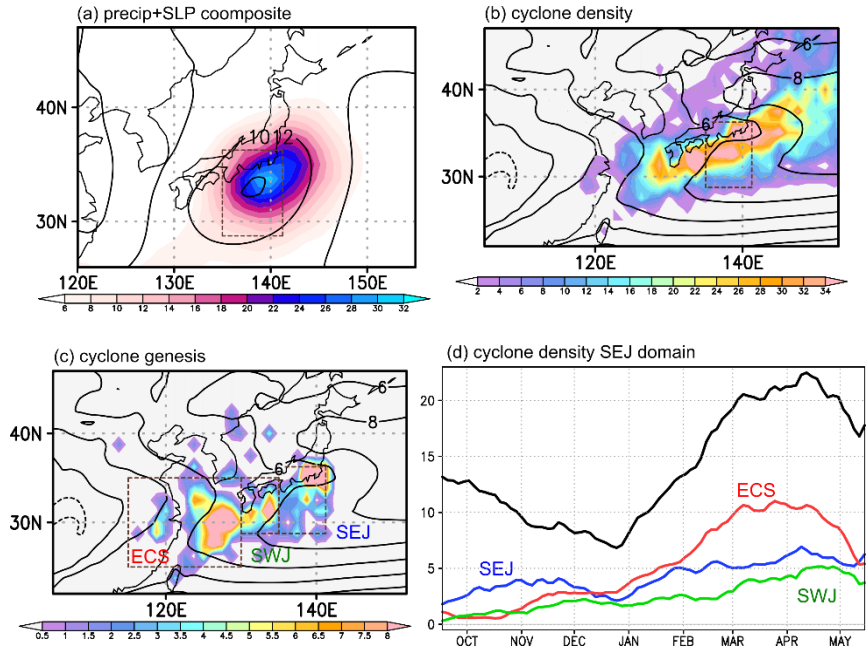

Fig. 3. (a) Springtime (31-day running-mean centered on the 19$^{th}$ March) composite map of precipitation (colors, mm/day) at time steps when one or more ETC centers are located over the SEJ domain [135°−141.25°E, 28.75°−36.25°N as indicated with dashed rectangles]. Contours indicate composited SLP (every 4hPa). (b) Climatological-mean density of ETC centers passing through the SEJ domain (color, 10$^{-4}$/6hr). Contours denote climatological-mean $U_{850}$ (every 2m/s). (c) Same as in (b), but for cyclogenesis occurrence (10$^{-4}$/6hr) for ETCs passing through the SEJ domain. (d) Seasonality of the climatological-mean ETC density (black line, 10$^{-4}$/6hr) averaged over the SEJ domain. Red, green, and blue lines signify the corresponding ETC density reconstructed only by ETCs generated within the ECS, SWJ, and SEJ domains indicated in (c).

Climatologically, cyclogenesis around the ECS (~30°N) peaks in spring north of the low-level jet axis (Fig. 4a). Nevertheless, the 850-hPa Eady growth rate (EGR; $\sigma = 0.31 \frac{f}{N} \frac{\partial U}{\partial z}$) (Eady 1949) around ~30°N peaks not in spring but in midwinter concomitantly with the maximized upper-level westerly jet intensity (Fig. 4b). Rather, the peak of the ECS cyclogenesis is more coherent with the maximized equatorward sea surface temperature (SST) gradient around ~30°N in late winter and spring (Fig. 4b). This seasonality of the ECS cyclogenesis contrasts with an early-winter peak of cyclogenesis occurrence over northern China around ~40°N, which is concurrent with the maximized low-level EGR. This suggests that factors other than climatological baroclinicity can be instrumental in forming the spring peak in cyclogenesis over the ECS and the Kuroshio cyclone density downstream.

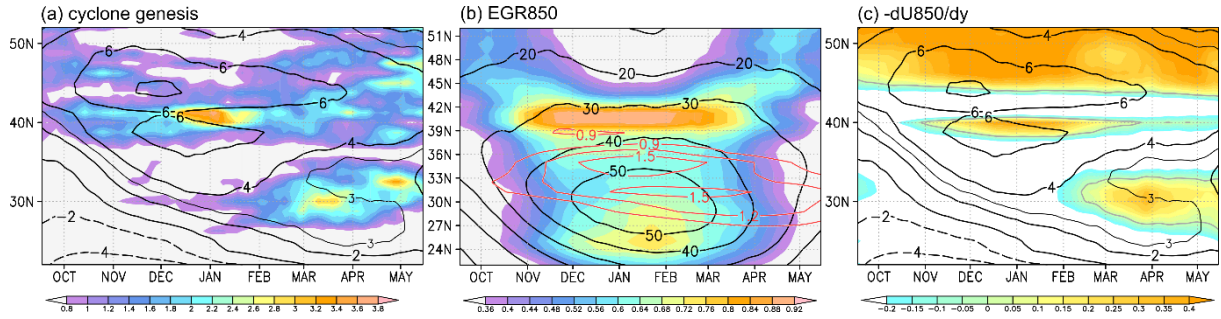

Fig. 4. Climatological seasonality in cyclogenesis occurrence (color, $10^{-4}$/6hr) averaged for 115°−130°E. Contours denote the corresponding seasonality of $U_{850}$. A tick mark on the abscissa in b-c represents the first day of a given calendar month. (b) Same as in (a), but for Eady growth rate (day$^{-1}$) at 850-hPa. Black contours denote the corresponding seasonality of $U_{300}$. Red contours denote seasonality of local $-dSST/dy$ (K/100km) based on brightness temperature over the ocean from the JRA-55. (c) Same as in (a), but for an equatorward shear of $U_{850}$ (m/s/100km).

## 4. Conditions for springtime ETCs and cyclogeneses around the ECS

To delineate mechanisms for the spring peak in cyclogenesis over the ECS, we investigate a typical condition for a cyclogenesis event within the ECS domain through a composite analysis. Figure 5a shows a composite map of unfiltered and high-pass-filtered SLP for genesis events within the ECS domain of ETCs that reach the SEJ domain. The unfiltered SLP field at the time step of the springtime cyclogenesis shows that a relatively weak composited cyclone is located over the ECS behind a strong anticyclone around Japan. The corresponding signatures of the ETC and anticyclone are also seen in the composited high-pass-filtered SLP field (Fig. 5a). The composited cyclone in unfiltered SLP is recognized as a weak pressure trough 6 hours before its genesis, and the corresponding weak cyclonic center is evident in the high-pass-filtered SLP (Fig. 5b).

Given the compatibility of the composited ETC between the unfiltered and high-pass-filtered fields (Figs. 5a-b), we investigate typical conditions for cyclogenesis within the ECS domain by decomposing a composited field $X$ into three components as

$X = X_C + X_L + X'$,

where subscripts C and L signify climatology and 8-day low-pass-filtered anomalous fields, respectively, and primes 8-day high-pass-filtered fields. In the following, we regard $X'$ as anomalies associated mainly with synoptic-scale ETCs, and $X_L$ as anomalous background fields for them. As shown in Fig. 5c, the high-pass-filtered geopotential height anomaly at

File generated with AMS Word template 2.0

850-hPa ($Z'_{850}$) composited for cyclogenesis events within the ECS domain is characterized by a cyclonic anomaly around the ECS and an anticyclonic anomaly downstream, corresponding well to the composited SLP' (Fig. 5a). The cyclonic $Z'$ around the ECS exhibits westward−tilting baroclinic structure, and its magnitude is strongest near the surface and decaying with height (Fig. 5d). This shallow cyclonic anomaly is consistent with Chang (2005). The upper-level cyclonic anomaly composited for the springtime cyclogenesis event in the ECS domain is weaker than its counterpart for midwinter events (not shown).

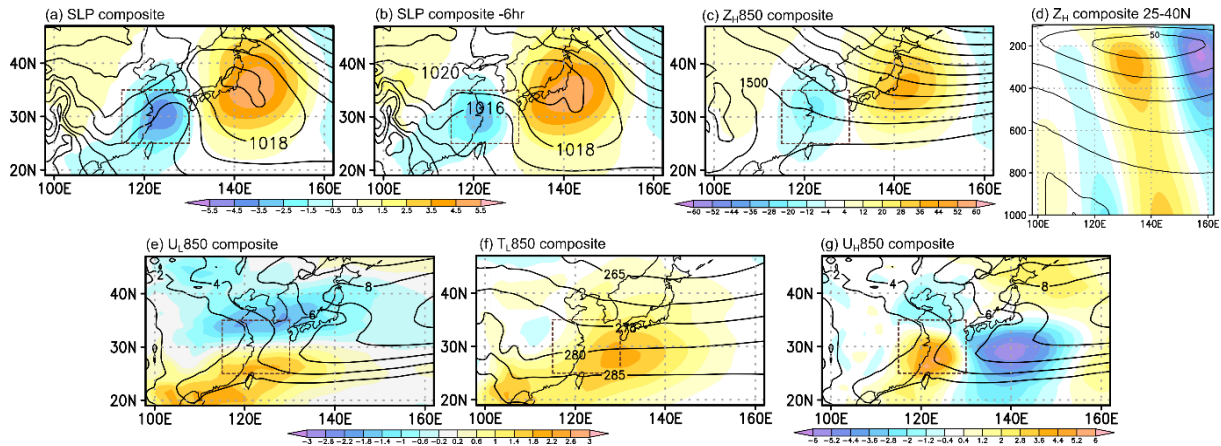

Fig. 5. (a) Springtime composite map of SLP (contours, every 2hPa) at time steps when one or more ETCs that pass through the SEJ domain [135°−141.25°E, 28.75°−36.25°N] in their lifetimes are generated over the ECS domain [115°−130°E, 25°−35°N; indicated with dashed rectangles]. Colors indicate composited SLP' (hPa). (b) Same as in (a), but for time steps 6 hours before the cyclogenesis events. (c) Same as in (a), but for $Z'_{850}$ (color, m). Contours denote $Z_{C850}$ (every 20m). (d) Zonal section of composited $Z'$ (color, m) averaged for 25°−40°N. Contours denote $U_C$ (every 10m/s). (e-g) Same as in (c), but for (e) $U_{L850}$ (m/s), (f) $T_{L850}$ (K), and (g) $U'_{850}$ (m/s). Contours denote $U_{C850}$ (every 2m/s) in (e, g) and $T_{850}$ (every 5K) in (f).SEJ

We further investigate low-level background anomalous conditions for the springtime ECS cyclogenesis events, in recognition of the strongest $Z'$ near the surface. The composited low-frequency westerly wind anomalies at 850-hPa ($U_{L850}$) are characterized by a zonally-elongated meridional dipole whose node is centered at ~30°N (Fig. 5e). The dipolar $U_{L850}$ anomalies act to enhance the cyclonic shear north of the low-level jet axis around the ECS by increasing and decreasing the westerly wind speed along the climatological low-level jet axis and a local $U_{C850}$ minimum north of it, respectively. The composited low-frequency temperature anomalies at 850-hPa ($T_{L850}$) exhibit overall warming over the subtropical and midlatitude western NP, which maximizes in the southeastern ECS and to the south of western Japan, acting to enhance climatological horizontal temperature gradient (mainly in the meridional direction) (Fig. 5f). The zonal scale of these low-frequency anomalies is substantially larger than that of high-frequency anomalies (Figs. 5c and 5g), indicative of their characteristics as "background conditions" for synoptic-scale ETCs. The above features can be seen also in cyclone-centered composite maps produced by shifting surface ETC centers horizontally in such a manner that they are all located at the origin of the coordinates (not shown).

The springtime cyclogenesis within the ECS domain is characterized also by a local increase in moisture (Fig. 6a), which is related to the free-tropospheric diabatic heating. This enhanced moisture is accompanied by composited poleward moisture flux south of ~30°N (Fig. 6a), in contrast to weak climatological-mean equatorward moisture flux around the ECS (Fig. 6b). The events of ECS cyclogenesis accompanies moisture increase also on the southern coast of China, where the lower troposphere is anomalously warm (Fig. 5e). The anomalous moisture flux along the southern coast of China is consistent with the anomalous low-level westerlies shown in Fig. 5e. Since the climatological-mean moisture gradually increases around the springtime ECS (Fig. 6c), the anomalous moistening in early- to mid-spring for cyclogenesis is likely to be regarded as a condition slightly ahead of season. The climatological-mean surface latent heat flux (LHF) around ~30°N, by contrast, maximizes in early winter (Fig. 6c), suggestive of the importance of the seasonality of the low-level circulation around the ECS.

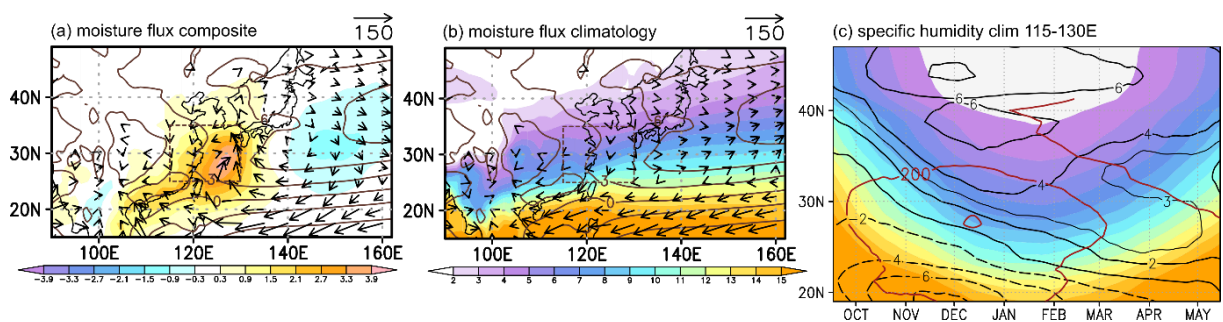

Fig. 6. (a) Same as in Fig. 5e, but for anomalous specific humidity from the climatology (color, mm/day) integrated vertically from the surface to 100-hPa. Arrows signify composited vertically-integrated moisture flux (g/kg m/s as indicated). Brown contours denote $U_{C850}$ (every 3m/s). (b) Same as in (a), but for the climatological-mean springtime vertically-integrated specific humidity (color, mm/day) and moisture flux (arrows, g/kg m/s as indicated). (c) Same as in Fig. 4a, but for climatological-mean vertically-integrated specific humidity (color, mm/day) and $U_{C850}$ (contours, m/s). Brown contours signify the corresponding climatological-mean upward surface LHF (W/m$^2$) averaged only over the ocean.

The moisture increase for the springtime cyclogenesis events around the ECS corresponds to distinct low-level diabatic heating ($Q$) slightly ahead of the composited ETC center (Fig. 7a), which acts to strengthen the ETC by lowering SLP around its center. The low-level diabatic heating around the ECS is due primarily to precipitation (convective + large-scale) (Fig. 7b), which is partially offset by weak diabatic cooling due to the other processes (vertical diffusion + radiative processes) (Fig. 7c). Similar offset can be seen also in southern China and the NP storm-track core region (Figs. 7b-c).

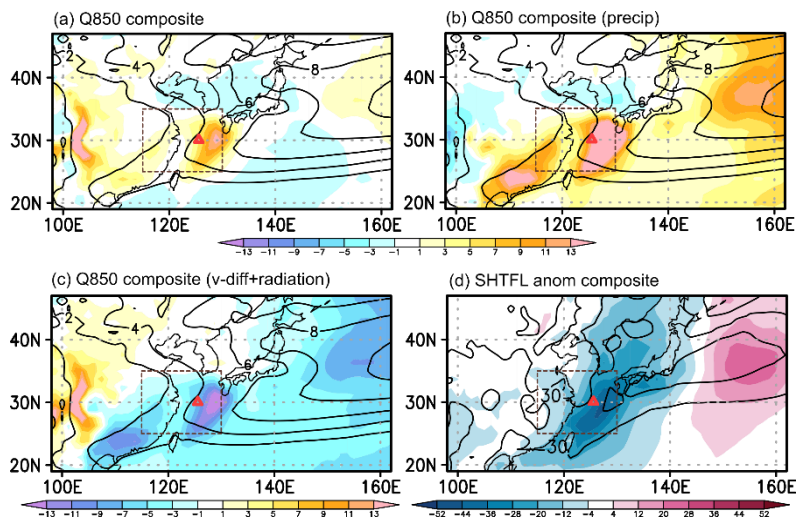

Fig. 7. (a) Same as in Fig. 5e, but for $Q_{850}$ (color, K/day). (b-c) Same as in (a), but for $Q_{850}$ due solely to (b) precipitation and to (c) vertical diffusion and radiative processes. (d) Same as in (a), but for composited anomalous upward surface SHF (color, W/m$^2$). Contours denote composited upward surface sensible heat flux (every 20W/m$^2$). Red triangles signify the composited surface cyclone center.

The composited springtime cyclogenesis around the ECS is accompanied by anomalous downward surface sensible heat flux (SHF, Fig. 7d), which is likely due to weaker near-surface wind and the anomalous low-level warmth in the corresponding composite (Fig. 5f). The downward SHF anomaly substantially weakens climatological-mean upward surface SHF around the ECS, which is likely contributing locally to the composited diabatic cooling associated with radiation and diffusion (Fig. 7c) and thereby acting to reduce the total diabatic heating (Fig. 7a). The reduction in the total diabatic heating suggests that local air-sea sensible heat exchange around the ECS is not an important factor for the springtime cyclogenesis around the ECS.

*Eddy energetics*

To obtain insights into the relative importance of relevant mechanisms for the lower-tropospheric springtime cyclogeneses in the ECS domain, we evaluate local eddy energetics (Okajima et al. 2022) based on the composited high- and low-pass-filtered fields. Energy conversion/generation rates for total eddy energy (sum of eddy kinetic and available potential energy) are defined as

$$CK = \frac{\overline{v'^2 - u'^2}}{2}\left(\frac{d\overline{u_L}}{dx} - \frac{d\overline{v_L}}{dy}\right) - \overline{u'v'}\left(\frac{d\overline{u_L}}{dy} + \frac{d\overline{v_L}}{dx}\right) = CK\text{x} + CK\text{y} \qquad (1a)$$

$$CP = \frac{R}{pS_p}\left(-\overline{u'T'}\frac{d\overline{T_L}}{dx} - \overline{v'T'}\frac{d\overline{T_L}}{dy}\right) \qquad (1b)$$

$$CQ = \frac{R}{pS_p}\overline{T'Q'} \qquad (1c)$$

where $S_p$ ($\equiv -T_C\, \partial \ln\theta_C/\partial p$) denotes a stability parameter and overbar a composited field. In (1), $u, v, T, Q$ and $R$ represent zonal and meridional wind components, temperature, temperature tendency due to diabatic processes, and the gas constant for dry air, respectively. CK, and CP denote barotropic and baroclinic energy conversion from the background state, respectively, and CQ signifies energy generation through diabatic processes. To focus on the low-level circulation anomaly at the cyclogenesis event (Fig. 5c), we compute the vertically-integrated energy conversion/generation rates from the surface to 700-hPa. Note that horizontal distributions of the energy conversion/generation rates at 850-hPa are qualitatively the same as the vertically-averaged rates from the surface to 700-hPa, except for CQ around the ECS (as shown later).

Around the composited ETC, positive CK acts to increase the eddy energy along the northern flank of the anomalously intensified background low-level jet (Fig. 8a), whose amplitude is larger than surrounding negative CK. The positive CK around the ECS, which is concentrated below the mid-troposphere (not shown), is due mainly to CKx in the entrance of the low-level jet (Fig. 8b). The contribution of CKy associated with the meridional shear of the jet is of secondary importance (Fig. 8c), although the maximum of the positive CKy is closer to the composited ETC center. Despite the northeastward tilt of the jet axis, the larger

File generated with AMS Word template 2.0

CKx than CKy suggests the efficient kinetic energy gain of meridionally-elongated eddies in the low-level jet entrance for cyclogenesis around the ECS. The cyclonic horizontal shear normal to the jet direction is likely to contribute to the formation of a cyclonic eddy through the conversion of shear vorticity into curvature vorticity (Bell and Keyser, 1993).

Around the composited cyclone, CP is positive and larger than CK (Fig. 8d), in association with warm advection by anomalous southerlies ahead of the cyclone. CQ is also larger than CK around the cyclone center (Fig. 8e), although negative to its southwest and northeast. These results indicate that a springtime relatively shallow ETC just generated around the ECS develops mainly through the baroclinic process, with secondary contributions of the barotropic and diabatic processes. At the 850-hPa, positive CQ is prominent around the ECS, which is largely canceled by negative CQ near the surface (not shown). Both the positive CQ and diabatic heating associated with precipitation (Fig. 7b) around the ETC center are in line with the importance of diabatic processes for the development of a cyclone around the ECS, as suggested in case studies by Takano (2002) and Ogura et al. (2005).

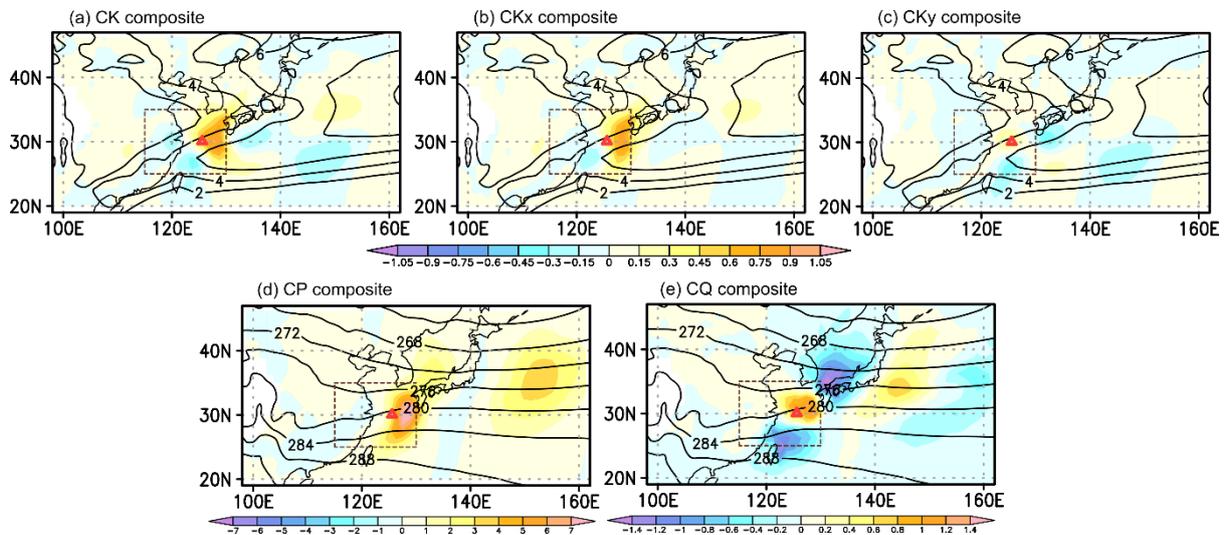

Fig. 8. (a-c) Same as in Fig. 5e, but for (a) CK, (b) CKx, (c) CKy, (d) CP, and (e) CQ (color, $10^{-4}$ J/s) integrated vertically from the surface to 700-hPa. Contours denote composited (a-c) $U_{L850}$ (every 2m/s) and (d-e) $T_{L850}$ (every 4K). Red triangles signify the composited surface cyclone center.

The amplitude of CK around the ECS 6 hours before the springtime cyclogenesis is relatively close to that at the cyclogenesis (Figs. 8a and 9a), despite the weaker composited ETC (Fig. 5b). Note that they are closer to each other at 850-hPa (not shown). Contrastingly, the amplitude of CP and CQ around the ECS is substantially smaller 6 hours before the cyclogenesis (Figs. 8d-e and 9a-b). These suggest the importance of the barotropic process for the initiation of the springtime cyclogenesis around the ECS.

The amplitude of CK around the ECS at the midwinter cyclogenesis is similar to that at the springtime cyclogenesis, consistent with the less distinct low-level jet around the ECS (Figs. 8a and 9d). By contrast, the amplitude of CP around the ECS at the midwinter cyclogenesis is greater than its counterpart at the springtime cyclogenesis (Figs. 8d and 9e). Negative CQ surrounding the positive CQ around the ETC center is lee prominent in midwinter than in spring (Figs. 8e and 9f). Together with the comparison between energetic terms at the springtime cyclogenesis and its preceding time step, these results indicate that the barotropic process is instrumental in forming the spring peak of the cyclogenesis occurrence around the ECS.

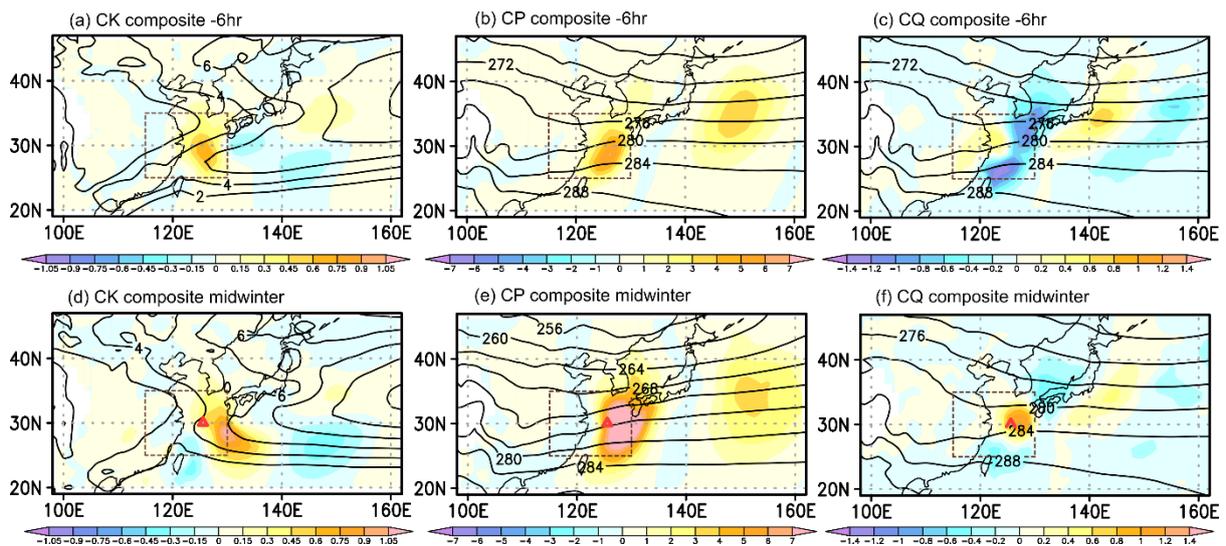

Fig. 9. (a-c) Same as in Figs. 8a, 8d, and 8e, respectively, but for the time steps 6 hours before the cyclogenesis events. (d-f) Same as in Figs. 8a, 8d, and 8e, respectively, but for the wintertime composite maps. In (d-f), red triangles signify the composited surface cyclone center.

*Atmospheric fronts*

The anomalous background conditions composited for the cyclogeneses around the springtime ECS, in which the low-level horizontal wind shear and temperature gradient are intensified (Fig. 5), suggest the importance of atmospheric fronts. This is compatible with Deveson et al. (2002), who pointed out that ETCs for which low-level processes are important tend to be located along atmospheric fronts. To assess the importance of atmospheric fronts, we objectively identify atmospheric fronts by applying two algorithms to the JRA-55 as by Soster and Parfitt (2022). One algorithm is based on the diagnostic formalized by Parfitt et al. (2017; hereafter P17), which focuses on low-level relative vorticity and horizontal temperature gradient. The other algorithm is based on the diagnostic used by Hewson (1998; hereafter H98), which focuses on equivalent potential temperature gradients. We assess the robustness of our results by comparing the two algorithms whose focuses are substantially different. We identify atmospheric fronts at 900-hPa as in P17. Additionally, a frontal grid point must have contiguous gridpoints for at least 500km to be identified as a front as by Schemm et al. (2015).

Climatological-mean frequency of atmospheric fronts in the ECS sector exhibits two pronounced maxima in winter (Figs. 10a-b). One is located at ~40°N peaking in early winter and the other is at ~24°N in mid- to late-winter by both algorithms. Those two latitudinal peaks are likely to be under the influence of land-sea thermal contrast across the coasts and topography, in addition to pronounced SST gradient in the southern portion of the ECS (Xie et al. 2002). On top of that, the frontal frequency around the ECS (~30°N) exhibits a spring peak coincides with the peak in the climatological-mean cyclonic shear (Fig. 4c) poleward of the low-level jet (Figs. 10a-b), which is more prominent through the P17 algorithm. The spring peak in the frontal frequency is likely related also to the spring peak in the meridional SST gradient (Fig. 4b), which is consistent with previous studies suggesting the importance of SST gradients for the frequency of atmospheric fronts (e.g., Parfitt et al. 2016). The frequent atmospheric fronts in spring are compatible with Utsumi et al. (2014), who showed through their analysis of weather charts that the frontal frequency is higher in MAM-mean than in DJF-mean.

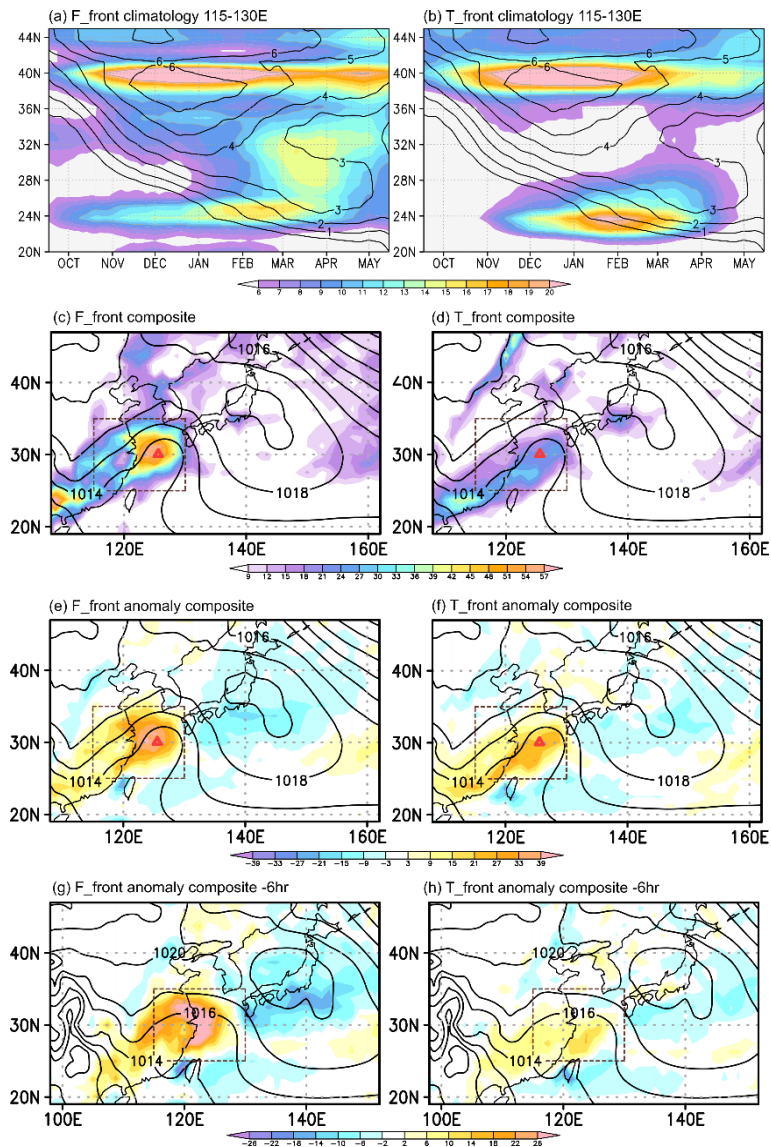

Fig. 10. (a, b) Same as in Fig. 4a, but for climatological-mean frontal occurrence (color, %) based on the algorithm by (a) P17 and (b) H98. Contours denote $U_{C850}$ (m/s). (c, d) Same as in Fig. 4e, but for climatological-mean frontal occurrence (color, %) based on the algorithm by (c) P17 and (d) H98. Contours denote composited SLP (every 2hPa). (e, f) Same as in (c, d), respectively, but for anomalous frontal occurrence (color, %). (g, h) Same as in (e, f), respectively, but for the time steps 6 hours before the cyclogenesis events. In (c-f), red triangles signify the composited surface cyclone center.

Both the P17 and H98 algorithms revealed that the frequency of atmospheric fronts tends to be markedly high around the ECS in the springtime cyclogenesis events (Figs. 10c-d). Those fronts tend to be observed in an area elongated in the SW-NE direction from southern China into the ECS. In fact, the composited anomalous frequency of atmospheric fronts increases substantially from southern China to the ECS and slightly decreases around Japan (Figs. 10e-f). The spatial distribution of the increase/decrease in the frontal frequency is similar between the P17 and H98 algorithms, which are influenced by the increased low-level cyclonic shear (Fig. 5e) and moisture (Fig. 6a), respectively. These results indicate that the springtime cyclogenesis around the ECS is characterized by the anomalously frequent atmospheric front in its vicinity on top of the climatologically high frontal frequency. The anomalously frequent atmospheric fronts around the ECS correspond to the increased horizontal wind shear and temperature gradient (Figs. 5e-f).

The anomalous increases in the frontal frequency are still evident in composite maps 6 hours before the cyclogenesis with some similarity between the two algorithms (Figs. 10g-h), though the composited SLP field only exhibits a pressure trough rather than a well-defined ETC. This suggests that the anomalously frequent atmospheric fronts related to the springtime ECS cyclogenesis may be regarded as background conditions for the genesis of synoptic-scale ETCs as shown in Fig. 5, although the composited frontal frequency can also be influenced by the composited synoptic-scale ETC itself. Similar to the climatological-mean frontal frequency, the composited anomalies in the frontal frequency are also larger in magnitude and more persistent through the P17 algorithm than through the H98 detection. This indicates that the low-level jet and associated horizontal wind shear around southern China and the ECS are fundamental components for the springtime cyclogenesis events around the ECS.

Our results thus point to the close relationship of atmospheric fronts with the springtime cyclogenesis events around the ECS, which is consistent with a case study by Takano (2002). Our results are also in line with Schemm et al. (2018), who showed that likelihood of "initial-front" cyclones is relatively high at ~30°N in EA and those fronts are mostly not attached to a parent ETC. Our results further indicate that the climatologically more distinct low-level jet and associated cyclonic wind shear in spring are important for the spring peak in the cyclogenesis occurrence around the ECS, given that both the seasonal enhancement into spring and fall-spring asymmetry are pronounced in the low-level jet and associated cyclonic shear (Fig. 4c) than in the low-level baroclinicity (Fig. 4b). Additionally, the close correspondence between the seasonality of cyclogenesis and frontal frequency suggests that the seasonality of the SST gradient around the ECS (Fig. 4b) is likely to influence the nearby low-level winds and near-surface atmospheric frontogenesis, contributing to the spring peak in the frequency of atmospheric fronts and cyclogenesis events around the ECS.

# 5. Mechanisms for the springtime development of the low-level jet around the ECS

This section focuses on the mechanisms for the climatological development of the low-level jet around southern China and the ECS, which is likely the key to the spring peak in the Kuroshio cyclone activity. Kim et al. (2013) suggested that a climatological increase in the thermal contrast between the continent and the ocean is concurrent with the development of the low-level jet around southern China. Here we hypothesize that the climatological seasonal evolution of the low-level jet around southern China and the ECS is explainable by that of diabatic heating. To test the hypothesis, we conducted a set of numerical experiments using a linear baroclinic model (LBM) developed by Watanabe and Kimoto (2000), focusing on differences between climatological-mean fields in January and FM (February−March)-mean. We carried out time integration with zonally-asymmetric basic states and steady diabatic heating as forcing taken from the JRA-55 reanalysis, with the T42 version of the LBM with 20 vertical levels. Specifically, for the climatological evolution between January and FM-mean (e.g., Fig. 11a), we conducted a set of experiments with the climatological difference in three-dimensional total diabatic heating rate between those calendar months (i.e., FM-mean−Jan) (e.g., colors in Fig. 11b) as a forcing and with the January climatology as a basic state (e.g., contours in Fig. 11b). We have confirmed that responses are not sensitive to the choice of a calendar month for a basic state (for instance, January or February; not shown). A response to a prescribed forcing is defined as the temporal average for the 31st-60th integration days. We confirmed that results are similar between differences for Feb−Jan and Mar−Feb, while the latter is stronger than the former (not shown).

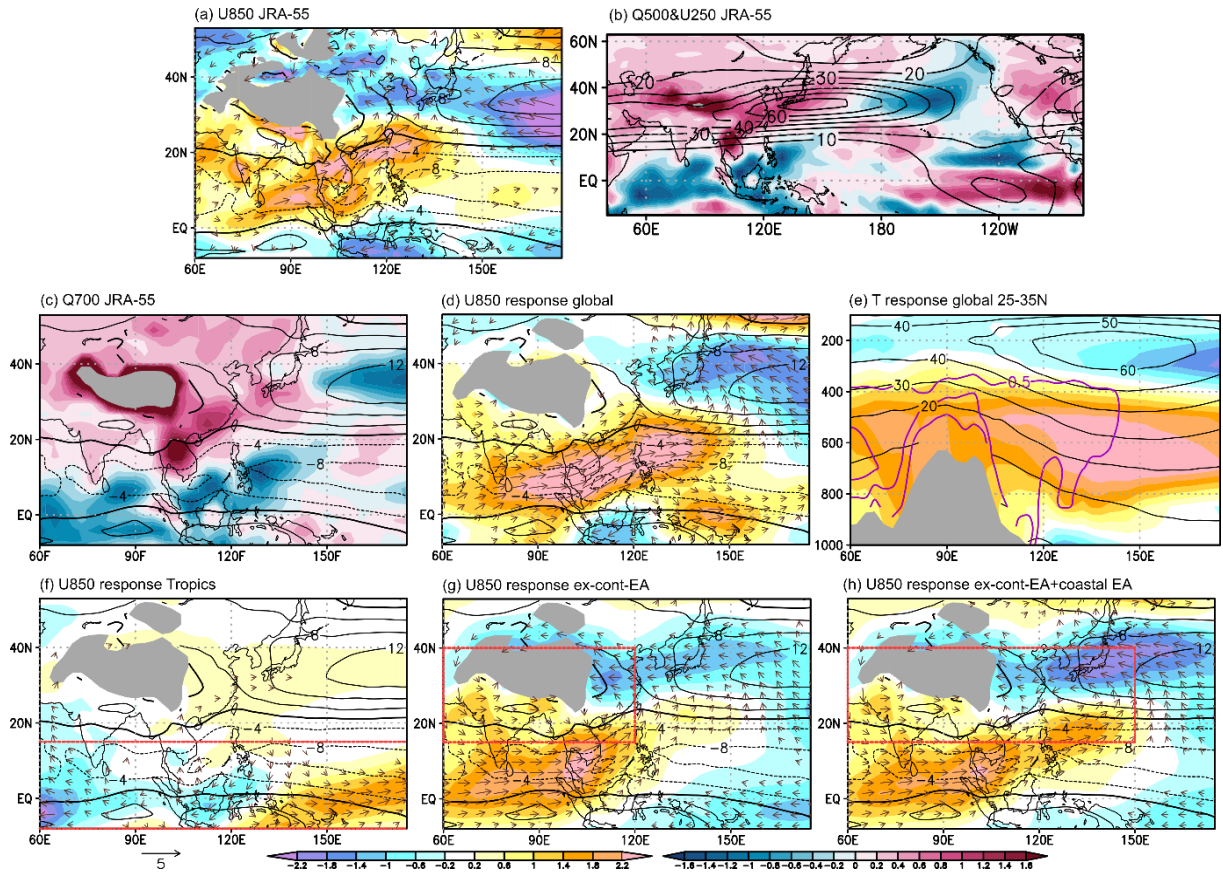

Fig. 11. (a) FM-mean−Jan difference in climatological-mean zonal wind velocity (color, m/s; positive and negative for westerly and easterly differences, respectively) and horizontal winds (arrows, m/s as indicated) at 850-hPa, based on the JRA-55 reanalysis. Contours denote climatological-mean $U_{850}$ (every 4m/s; positive for westerlies, bold lines for zero) for January. (b) Same as in (a), but for $Q_{500}$ difference (color, K/day). Contours denote climatological-mean $U_{250}$ (every 10m/s; positive and negative for westerly and easterly responses, respectively) in January. (c) Same as in (a), but for $Q_{700}$ (color, K/day). (d) Response of westerly wind speed (color, m/s) and horizontal winds (arrows, m/s as indicated) at 850-hPa to the FM-mean−Jan difference in diabatic heating. (e) Longitudinal section of the LBM response in temperature (color, K) averaged for 25°−35°N. Purple contours signify the corresponding $Q$ forcing (+0.5, 1.0, and 1.5K/day). Black contours denote corresponding climatological-mean $U$ (every 10m/s) for January. (f-h) Same as in (d), but for the response to the difference in diabatic heating only in the respective domains of (f) the entire Tropics [0°−360°E, 15°S−15°N], (g) extended continental EA [60°−120°E, 15°−40°N], and (h) extended continental and coastal EA [60°−150°E, 15°−40°N] as indicated by dashed red rectangles.

In the Tropics, the climatological evolution of diabatic heating from January to FM-mean is characterized by weakened (strengthened) heating north (south) of the equator over the central and eastern Pacific, in addition to reduced heating in the equatorial Indian Ocean and Maritime Continent (Fig. 11b). In addition, the climatological diabatic heating around EA, especially over the Tibetan Plateau, Indochina Peninsula, and southern China, becomes stronger from January to FM-mean (Fig. 11c). They are likely related to a seasonal increase in insolation. Those centers of the climatological increase in diabatic heating around EA are close to the climatological-mean axis of the subtropical jet (Fig. 11b).

The LBM response to the climatological evolution of diabatic heating from January to FM-mean shown in Fig. 11d qualitatively reproduces the corresponding climatological enhancement of the low-level jet and associated horizontal wind shear observed around EA (Figs. 11a), though the westerly and easterly responses in the Tropics and NP, respectively, are somewhat overestimated. Around the ECS, a low-level cyclonic response around ~30°N acts to strengthen the background cyclonic shear (Fig. 11d), which is accompanied by an overlying warm response peaking around the 600~700-hPa levels downstream of the Tibetan Plateau (Fig. 11e). These results of the LBM experiment indicate that the climatological evolution of the low-level jet and associated horizontal wind shear around the ECS from January to FM-mean can be interpreted largely as a linear response to the seasonal evolution of diabatic heating. Note that the LBM response does not include feedback forcing by eddies, which is likely to explain the difference between the LBM response and the observed seasonal evolution.

We further assess the relative importance of the seasonal evolution of diabatic heating in the Tropics and extratropics. We conducted similar LBM experiments but with the seasonal evolution of diabatic heating only within a selected domain under the same basic state as the global forcing experiment. A linear response solely to the Jan−FM-mean evolution of diabatic heating in the Tropics [0°−360°E, 15°S−15°N] is found very weak (Fig. 11f). By contrast, the corresponding response solely to the seasonally evolving diabatic heating in an extended continental EA region [60°−120°E, 15°−40°N] more strongly reinforces the background cyclonic shear around the ECS (Fig. 11g), though somewhat underestimating the response to the global forcing (Fig. 11d). The diabatic heating responsible for the response is attributable mainly to precipitation and secondarily to radiative processes (not shown). Interestingly, by expanding the extended continental EA region further eastward to [60°−150°E, 15°−40°N] to include coastal EA, the cyclonic response around the ECS becomes even stronger and more comparable in magnitude to the response to the global forcing (Fig. 11h). This suggests that there may be a positive feedback loop in which the seasonal enhancement of diabatic heating around the ECS, which is likely associated with the seasonal increase in cyclone/cyclogenesis occurrence, acts to further strengthen the low-level jet and associated cyclonic shear around the ECS.

These results suggest that the seasonal evolution of diabatic heating in EA plays a pivotal role in the climatological springtime enhancement of the cyclogenesis occurrence around the ECS, and consequently, Kuroshio cyclone activity. Our finding is consistent with the climatologically increased thermal contrast between the continent and the ocean concurrently with the development of the low-level jet around southern China (Kim et al. 2013). This is also compatible with Lee et al. (2013), who pointed out that the Tibetan Plateau is important for the midwinter minimum of transient eddy activity in EA. We indeed obtained similar strengthening of the cyclonic shear around the ECS in response to diabatic heating forcings only around the Tibetan Plateau [70°−110°E, 22°−40°N], though somewhat weaker (not shown).

## 6. Summary and discussions

The present study investigates the mechanisms for the spring peak in the frequency of ETCs in EA that typically move along the Kuroshio (i.e., Kuroshio cyclones). Figure 12 schematically illustrates factors relevant to the spring peak in the Kuroshio cyclone activity. In spring, the climatological diabatic heating around EA, especially around the Tibetan Plateau, Indochina Peninsula, and southern China, becomes stronger than in winter, leading to a mid-tropospheric warm response downstream along the climatological subtropical jet. The warm response accompanies a cyclonic response around EA that acts to strengthen the low-level jet in southern China and the ECS. Climatologically strengthened cyclonic shear north of the low-level jet axis and frequent atmospheric frontogenesis serve as favorable background conditions for surface cyclogenesis around the ECS. Those ETCs generated around the ECS develop through the baroclinic and diabatic processes fueled by strengthened poleward moisture flux with a secondary contribution of the barotropic process. These ETCs move eastward to form the spring peak in ETC density along the Kuroshio.

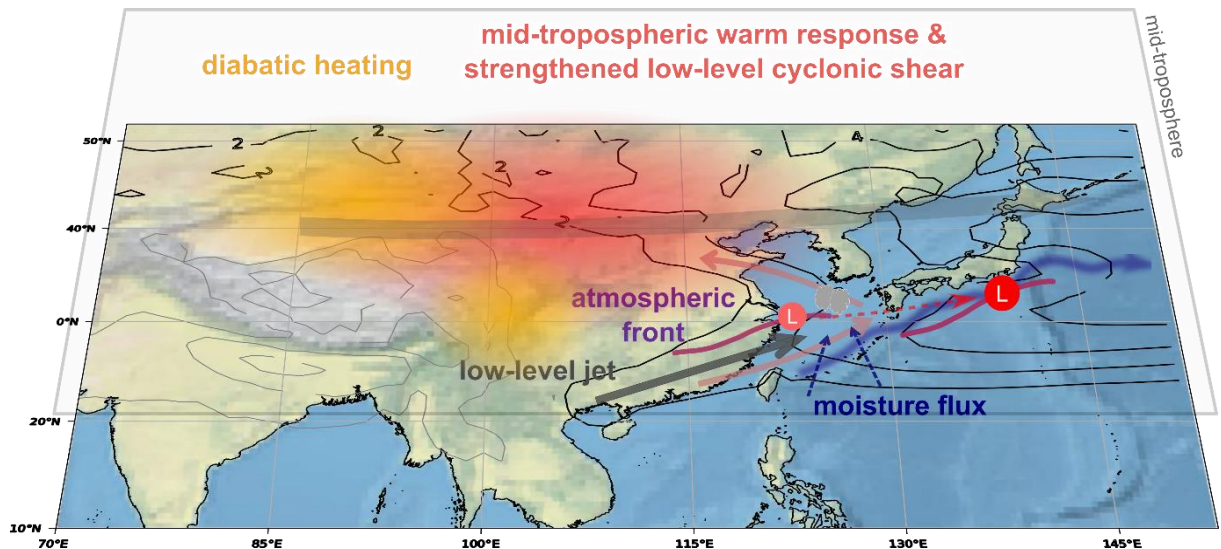

Fig. 12. Schematic for the mechanisms for the spring peak in the Kuroshio cyclone frequency in EA. Black and grey contours denote climatological $U_{850}$ (m/s) in March.

The LBM experiments suggests the importance of the Tibetan Plateau for the seasonal evolution of the jet stream and climate system in EA, which has been intensively studied and pointed out in the context of the onset of EA summer monsoon (e.g., Wu and Zhang 1998; Huang et al. 2023). The results in this study suggest that the Tibetan Plateau is important also for the seasonal evolution of the climate system in EA from winter to spring, which is related to the spring peak in the cyclone activity and climatological precipitation, as well as in the pre-monsoon rainy season in southern China. Our findings are consistent with LinHo et al. (2008), who pointed out the importance of the gradual weakening of the land-seas thermal contrast due to the increase of insolation for the spring rainy season in southern China. They also suggested that passage of the dry phase of the Madden-Julian Oscillation (MJO) over the Maritime Continent around mid-March (Wang et al. 2024), which is coincident with the termination of the Indonesian–Australian summer monsoon, leads to the rainfall increase over southern China through rapid weakening of the cross-equatorial Hadley cell. Our LBM experiments show that the seasonal evolution of convective activity in the Tropics is unlikely to be instrumental in the spring peak in ETC activity in EA. Nevertheless, the potential influence of MJO events is worth investigating. Additionally, the role of the anticyclonic anomaly and associated wave train downstream of cyclogenesis events around the ECS is to be studied.

24File generated with AMS Word template 2.0

The results shown in this study are based on a single tracking algorithm. Statistics regarding feature tracking are rather sensitive to the tracking algorithm, the parameter being tracked, as well as whether any filtering has been applied in advance (Neu et al. 2013), though the winter-mean climatological density of cyclones by the algorithm used in this study is compatible with previous studies (Okajima et al. 2023).

The activity of wintertime Kuroshio cyclones moving off the south coast of Japan tends to increase during El Niño (Ueda et al. 2017). Interannual variability of the springtime ETC activity in EA and its potential predictability arising from SST variability need to be studied. ETCs generated around the ECS and moving along the south coast of Japan are likely to be influenced also locally by SST over the ECS, Kuroshio, and KE (Nakamura et al. 2012; Hayasaki et al. 2013). Thus impacts of SST resolution (e.g., Masunaga et al. 2018) on Kuroshio cyclones should be evaluated in our future study. The importance of the Tibetan Plateau warrants future studies on the ability of climate models to represent the springtime Kuroshio cyclone activity and its future projections, including their reproducibility of the conditions around the Tibetan Plateau (e.g., Portal et al. 2023). This would help us understand how the seasonal increase in climatological diabatic heating in EA takes place.

*Acknowledgments.*

The authors thank Frederick Soster for his assistance with the frontal data. This study is supported in part by the Japanese Ministry of Education, Culture, Sports, Science and Technology (MEXT) through the Arctic Challenge for Sustainability II (ArCS-II; JPMXD1420318865) and the Advanced Studies of Climate Change Projection (SENTAN; JPMXD0722680395, by the Japan Science and Technology Agency through COI-NEXT JPMJPF2013, by the Japanese Ministry of Environment through Environmental Restoration and Conservation Agency Fund JPMEERF20222002, and by the Japan Society for the Promotion of Science (JSPS) through Grants-in-Aid for Scientific Research 19H05702 (on Innovative Areas 6102), 20H01970, 21H01164, 22H01292, and 22K14097. RP acknowledges support from NOAA Climate Variability and Predictability Program NA22OAR4310617.

*Data Availability Statement.*

File generated with AMS Word template 2.0

The JRA-55 reanalysis is from the Data Integration and Analysis System (DIAS; https://search.diasjp.net/en/dataset/JRA55). The monthly climatologies in Fig. 1 are from JMA's ClimatView (https://ds.data.jma.go.jp/tcc/tcc/products/climate/climatview/frame.php). The GPCP v3.2 monthly precipitation data is from https://disc.gsfc.nasa.gov/datasets/GPCPMON_3.2/summary. Figure 13 is produced with Natural Earth (https://www.naturalearthdata.com/).

File generated with AMS Word template 2.0